\documentclass[12pt]{article}

\begin{document}

\title{Is Science going through a critical stage?}

\author{Luigi Foschini\\
{\small CNR - Institute FISBAT, Via Gobetti 101, I-40129, Bologna 
(Italy)}\\
{\small E-mail: L.Foschini@fisbat.bo.cnr.it}}

\maketitle

\begin{abstract}
The unexpected discoveries at the beginning of the century, particularly thanks to
Heisenberg, Bohr, and G\"{o}del, has driven the science to drastic changes, opening
new, extraordinary, and infinite research fields. After this, many scientists saw, and
still today see, a crisis, with dreadful meaning, in the science. However, this crisis
is only present in that type of science, driven by determinism, which is strictly
linked to the common sense.
\end{abstract}

\vskip 24pt
\begin{center}
Originally published in italian:\\
L. Foschini, La Scienza \`{e} in crisi?, 
\emph{AEI} \textbf{83}, (1996), 455-458.\\
English translation by Ms. Simona Baldoni. Revised by the author.  
\end{center}

\vskip 24pt

The first half of this century has been one of the most intense period 
in human history: tragic episodes - the two world wars - were 
interwoven with moments of great cultural activity.  At the end of the 
XIX century, the world was still permeated with an extremely 
determinist spirit, which was the result of the philosophical ideas of 
the previous centuries: Enlightment first and Positivism later on.  
The former - which historically coincides with the XVIII century - 
tried to ``enlighten'' the mind of man, seized by ignorance and 
superstition, through science and knowledge.  The latter - borne on 
the wave of the great scientific discoveries of the time - may well be 
seen as an evolution of Enlightment.  The word `positivism' has been 
introduced by Auguste Comte to make a distinction among the scientific 
stage of knowledge man had reached and the metaphysical and the 
theological ones.  Positivism recognizes in science the only and real 
knowledge and takes it as a model for every other part of knowledege.  
Its scientific rationality and experimental methodology became 
universal, stonewalling any other kind of reason.  This kind of 
rationality was based upon the convinction that only facts have value, 
along with their prevision and understading: this was the only way to 
reach a positive control of phenomena, according to the needs of 
mankind.  Materialism and Scientism came later on and brought to their 
extremes the positivistic ideas on the superiority of science, of 
evolutionism and denied the existance of any kind of metaphysics.
  
However, towards the end of the XIX century, the impossibility in 
obtaining a satisfactory explanation of some phenomena showed the deep 
deficiencies of Positivists' deterministic materialism{\footnote{For a 
throughout tractation of the history of physics of this period see 
\cite{SEGRE}, \cite{BELLONE}, \cite{DAUMAS}, \cite{RHODES}.}}.  These 
problems were overcome thanks to the introduction of a mathematical 
formalism, the first step towards quantum mechanics.  In 1925, the 
German Werner Heisenberg, Max Born and Pascual Jordan provided a first 
formulation based on the matrix calculus.  One year later, Erwin 
Schr\"{o}dinger proposed an alternative way, using waves.  Therefore, 
in 1926, two formalisms were availaible, the wave mechanics and the 
matrix, capable of investigating on atomic phenomena.  The 
symbolic formulation by Paul A. M. Dirac, an English physicist was 
then added to these.

The three theories proved to be equivalent; perhaps the most important 
consequence was to discover that both light and matter have 
particle--like and wave--like properties.  The old debate on the 
nature of light, dating back to Newton and Huygens, seemed therefore 
solved, though a new problem arose: that of how could light be both a 
particle and a wave.  Werner Heisenberg and Niels Bohr showed that 
this inconsistency was only apparent: in 1927, the former published a 
legendary article, where he enounced the famous principle of 
indeterminacy \cite{HEISENBERG}.  This principle can be summarized 
with this definition: it is not possible to know simultaneously and 
with precision the position and the moment of a particle.

On 16th September of the same year, during an intenational congress at 
Como, held for the centenary of Alessandro Volta's death, Niels Bohr 
introduced the principle of complementarity \cite{BOHR1} of which 
there is no precise enunciation.  Bohr spoke about it in this way:

\begin{quotation}
	The very nature of the quantum theory thus forces us to regard the 
	space-time co-ordination and the claim of causality, the union of 
	which characterizes the classical theories, as complementary but 
	exclusive features of the description, symbolizing the 
	idealization of observation and definition respectively 
	\cite{BOHR1}.
\end{quotation}

Along with this Bohr brought some examples, such as the debate on the 
nature of light and of the ultimate parts of matter; after all, what 
this principle enounces is that matter has a dual behaviour, 
wave--like and particle--like.

One of the main consequences of these principles regards the concept 
of reality.  On this subject Bohr writes \cite{BOHR1} that each 
observation of atomic phenomena involves a considerable interaction 
with the measure instrument: therefore neither the phenomenon, nor the 
instrument can be assigned an indipendent and objective reality, in the 
ordinary sense of physics.  On the other hand, the great Danish 
physicist adds that an element of arbitrariness is already implicit in 
the concept of observation, for it depends on who is considered to be 
the observator and who is the observed.

The fact that classical physics could develop, and the reason why it is 
not necessary to throw it away today, is due to the extremely small 
value of the action quantum (we remind that $h=6.62618\cdot 10^{-34}$ 
Js) as to the actions playing in the common sensorial perception.  
However, these principles should always be borne in mind when speaking 
about any sector of physics, and even of science, to remember that what 
one is speaking about is not nature and not even its image.

It is Niels Bohr again who provides us with the correct definition: 
physics regards what we can \emph{say} about nature, it is the writing 
of evidences around a praxis (see \cite{PETERSEN}).  Bohr gave this 
definition of the experiment \cite{BOHR2}:

\begin{quotation}
\ldots with the word ``experiment'' we can only mean a procedure regarding
which we are able to communicate to others what we have done and what we have learnt.
\end{quotation}

We have to remember that every time one formulates a theory, one sets 
hypotheses of effectiveness which generally consist in excluding one 
factor or another.  How many times one has supposed that a phenomenon 
was linear?  How many times one has supposed it ideal (rigid bodies, 
geometrical bodies, material points).  How many times one supposes 
that the resistance of an electric device is negligible?  How many 
times is friction considered negligible?  And taking into 
consideration the two-bodies problem, one forgets the 
interactions among the bodies of the universe, isn't it?  Physics and 
engineering are permeated with hypothesis of this kind, without which 
we could not adventure in building models or formulating theories.  
The more or less indirect consequences for engineering are constituted 
by the introduction of the safety factor, by the concept of 
reliability of devices; in physics we speak about the experimental 
errors, the domain of validity of a theory and so on.  With all these 
hypotheses, how could one say \emph{what is the nature}?  This is not 
a mere philosophical speculation, a sophism, a formal problem.

Words have a fundamental importance in all human activities.  In 1623, 
Galileo Galilei wrote in his book \emph{Il Saggiatore} that nature was 
like a book written in a mathematical language \cite{GALILEI}.  
Quantum physics shows that mathematics is not the language of nature, 
but an invented language created by man through which it is possible 
to say something about nature.  It is a refinement (or an 
impoverishment?)  of language eligible to represent those relations 
for which the commmon word would result imprecise or far too complex.  
The importance of mathematics as a language for physics has come up 
with the advent of quantum physics indeed.  As a matter of facts, 
contemporary physics can be briefly divided in two parts: one based on 
the analysis of the phenomena for which we have direct experience, of 
the everyday world and that express itself with the common language.  
The other is consituted by those phenomena which regard the extremely 
small (quantum physics) and the speeds near to that of light (relativistic physics), 
of which we have neither direct experience nor an adequate language to describe 
them, except that of mathematics.  When speaking about quantum and 
relativistic physics an uncorrect use of words could lead to 
misunderstandings creating inappropriate images.  Heisenberg's comment 
is that we have then to resign to the fact that the experimental 
observations in the extremely small and in the extremely wide cannot 
give more than an intuitive image; whithin these fields we have to 
learn to do without intuition \cite{HEISENBERG2}.

However, one is not allowed to think that mathematics is the last hope 
for Determinism.  As a matter of fact the analogous of the principles 
of indeterminacy for mathematics was expressed by Kurt G\"{o}del in 
1931 \cite{GODEL}.  In his article, he stated the impossibility to realize 
the hilbertian program: in 1900, during the \emph{Second 
International Congress of Mathematicians} in Paris, David Hilbert 
introduced a list of 23 problems which covered the most different 
fields of mathematics \cite{HILBERT}.  Among these, point 2, relative 
to the demonstration of non-contradiction of arithmetics, deserves a 
particular attention.  From Hilbert's viewpoint all mathematical 
theories should have been reduced to formal systems: then this would 
have been enough to demonstrate the non-contradiction.  In 1930, 
G\"{o}del wrote an article, which was published one year later where 
he demonstrated that this was not possible.  As a matter of facts, 
within a sytem like that expressed by Bertrand Russell and Alfred N. 
Whitehead in the \emph{Principia Mathematica} it is possible to 
express propositions which are not decidable within the system's 
axioms.  One can view this as the impossibility of defining each 
concept through a unique and defined linguistic universe.

The expression of these formulations made many scientists feel 
dejected, they cried science had reached its end.  Many of them did 
not want to leave the anchor of an objective reality, existing 
independently from everything else: Einstein, Planck, Schr\"{o}dinger 
so to cite some of them.  Einstein's positions has nearly become 
legendary: during the Solvay congress, held in 1927 in Bruxelles, he 
expressed many ideal experiments which should have invalidated the 
principles of quantum mechanics, though they were punctually disproved 
by Bohr and Heisenberg.  Einstein recognized the validity 
of quantum mechanics though he refused to accept it, summarizing this 
aprioristic refusal with the famous sentence ``God does not play 
dice'' to which Bohr replied that ``Our problem does not not consist 
in telling God how he has to govern the world'' (see 
\cite{HEISENBERG2}, Chapter 7).

Nowadays, many physicists still refuse quantum physics for they 
consider it irrational or mystical.  Even today they are calling it 
crisis: Marcello Cini speaks about a ``paradise lost'' \cite{CINI} and 
not long ago Eugenio Sarti wrote an article for this Review 
\cite{SARTI}.
  
Using the term `crisis' they suggest something dreadful, that will lead 
to the very end of science.  Some scientists think that this crisis is 
already operating and it is the result of the principles up to now 
discussed, others think it will come along with the Great Unified 
Theory.  Nevertheless, the word `crisis' shows no dreadful meanings: it 
derives from the Greek $\kappa\rho\iota\sigma\iota\varsigma$, which in 
turn is linked to $\kappa\rho\iota\nu\omega$, which means `to 
divide' and metaphorically `to decide' {\footnote{It is interesting 
to see that the word `science' as well comes from the sanscrit root 
\emph{skad-}, \emph{skid-}, which means `to cut', `to mince'; knowledge 
should then result from the separation of notions.}}.  It were the 
Greeks the first to introduce the process of analysis as a 
\emph{division} of a thesis in propositions, leading more easily to 
truth.  If, within a theory, we separate or, better, underline, some 
essential laws we could then consider them as principles for a new 
theory.  Analysis is essential to science because, as Paolo Zellini 
writes \cite{ZELLINI1}, this \emph{divide et impera} preserves the 
thesis from the risk of a total rejection deriving from an excess of 
rigidity or from an unconscious desire for barreness; it diverts the 
attention on the central parts of the demonstration, revealing those 
hidden hypotesis faked by counterexamples.  Counterexamples do not 
actually fake the thesis as a whole, rather some of its implicit 
assumptions, not always recognized when trying to demonstrate it.  A 
scientific theory thrives until these implicit assumptions, eligible 
to be faked by counterexamples, exist.  The surprising thing is that, 
as Imre Lakatos writes (see \cite{ZELLINI1}), these concepts grow by 
themselves and generate a maze of problems.

The word, let it be mathematical or from the common language, cannot 
be managed: all those huge programs that claim to ground some 
discipline, that believe to find the ultimate formula, the pill, the 
magic potion which gives knowldge, well all these programs cannot but 
result in some grounds crisis: as Lakatos wrote (see \cite{ZELLINI1}), 
both certainty and banality are infantile illnesses of knowledge.

However, this ground crisis should not degeneratre in a skeptical 
cynism or in mysticism.  The principles up to now discussed suggest 
the solution for they do not imply the end of science, not at all.  
That is the good of it, because thanks to these principles an unknown, 
endless universe is opened wide to us, waiting to be analyzed.

The infinite that we still consider with mistrust, in ancient Greece 
was $\alpha\pi\epsilon\iota\rho o\nu$, `without limits', `unlimited'.  
Aristotle (see \cite{ZELLINI2}) wrote that infinite is not that out of 
which there is nothing, though that out of which there is always something. 
The unlimited cannot therefore be regaded as a complete whole: what is 
completed has an end and the end is a limiting element, while 
$\alpha\pi\epsilon\iota\rho o\nu$ shows for its very meaning the 
absence of any limits (see \cite{ZELLINI2}).  Tannery (see \cite{ZELLINI2}) 
suggested to derive Anassimander's $\alpha\pi\epsilon\iota\rho o\nu$ from 
$\pi\epsilon\iota\rho\alpha$ meaning `knowledge', instead of deriving it 
from $\pi\epsilon\rho\alpha\varsigma$, `limit': in this way the 
unlimited became the unknowable (see \cite{ZELLINI2}).

Anyway, the importance on the infinite is in its becoming.  As 
Aristotle wrote in his \emph{Physics}, if the limit is what make each 
object exist, giving it a form, the infinite is its opposite principle 
that prevents each object to be fixed in its limits, within its 
boundaries.  Anassimander thought that the $\alpha\pi\epsilon\iota\rho 
o\nu$ was the very principle of becoming.

When somebody will find the ultimate formula, then the end of 
mathematics and physics will come. Bohr's, G\"{o}del's and 
Heisenberg's essays do not state the limits of science, on the contrary 
they show the non-existance of these boundaries.

These are no comfortable ways, as some of the stronger supporter of 
Determinism thought, rather these are big steps forward made by 
science.  It is not by chance that this revolution has taken place in 
mathematics, a highly sophisticated language, and in physics, that uses 
this language intensively.

If we really look for a crisis in science we will see that there is 
one in mathematics and in physics for these sciences look for 
completeness, for the seek their beginning to determine their end 
though this was already clear in 1927 for physics and in 1931 for 
mathematics.  Nevertheless there is who prefers to ignore these 
questions and go on trying to build a deterministic world, complete, 
limited, foreseable.  We are not stating that this is wrong or that 
this should not be done, for when one researches it is not possible to 
say what is to be done or not.  The important is to do and then who 
knows if from these researches some interesting hint may come.

However, it is essential for research to investigate the infinite.

\end{document}